\documentclass[%
 reprint,
 superscriptaddress,
 amsmath,amssymb,
 aps,
 pra,
]{revtex4-2}

\usepackage{physics}
\usepackage{graphicx}
\usepackage{subfigure}
\usepackage{dcolumn}
\usepackage{bm}
\usepackage[colorlinks,
            urlcolor=blue,
            linkcolor=blue,
            anchorcolor=blue,
            citecolor=green
            ]{hyperref}

\usepackage{pifont}

\newcommand{\vect}[1]{\mathbf{#1}}
\newcommand{\Eq}[1]{Eq.~\eqref{#1}}

\newcommand{\Fig}[1]{Fig.~\ref{#1}}
\newcommand{\beq}{\begin{equation}}
\newcommand{\eeq}{\end{equation}}

\begin{document}

\preprint{APS/123-QED}

\title{Universal Energy Functionals for Trapped Fermi Gases in Low Dimensions}

\author{Jiansen Zhang}
 \email{jiansenzhang\_prc@stu.pku.edu.cn}
\author{Shina Tan}%
 \email{shinatan@pku.edu.cn}
\affiliation{International Center for Quantum Materials, Peking University, Beijing 100871, China}%

\date{\today}

\begin{abstract}
We study the system of trapped two-component Fermi gases with zero-range interaction in two dimensions (2D) or one dimension (1D). We calculate the one-particle density matrices of these systems at small displacements, from which we show that the $N$-body energies are linear
functionals of the occupation probabilities of single-particle energy eigenstates. A universal energy functional was first derived in 2011 \cite{Tan2011} for trapped zero-range interacting two-component Fermi gases in three dimensions (3D). 
We also calculate the asymptotic behaviors of the occupation probabilities of single-particle energy eigenstates at high energies. 
\end{abstract}

\maketitle


\section{\label{sec:introduction}Introduction}

The zero-range interacting systems are good models for many physical systems, such as the ultracold Bose gases \cite{zero-range-bose1, zero-range-bose2, zero-range-bose3, zero-range-bose4, zero-range-bose5}, ultracold Fermi gases \cite{zero-range-fermi1, zero-range-fermi2, zero-range-fermi3, zero-range-fermi4}, 
and few-nucleon systems \cite{zero-range-nuclei1, zero-range-nuclei2, zero-range-nuclei3}.
If the mean inter-particle distance $d$ and the thermal de Brogile wavelength $\lambda$ are both much larger than the range $r_e$ of the interaction
between the particles, the system may be approximated as a zero-range interacting system, and it has
universal properties that do not depend on the details of the interaction. These universal properties depend on the interaction potential through the $s$-wave scattering length $a$, which characterizes the low-energy scattering properties. This universality exists in the Bose systems \cite{universality-bose1, universality-bose2, universality-bose3, universality-bose4, universality-bose5, universality-bose6}, the Fermi systems \cite{universality-fermi1, universality-fermi2, universality-fermi3, universality-fermi4}, and the mixtures \cite{universality-mixture1, universality-mixture2, universality-mixture3, universality-mixture4}.

For the 3D two-component Fermi systems with $s$-wave contact interaction,
it was found that there exists a universal parameter $\mathcal{I}_{3\mathrm D}$, called contact, characterizing the tail of the momentum distribution at large $\vect k$, where $\hbar\vect k$ is the single-particle momentum,
and $\hbar$ is Planck's constant over $2\pi$,
and that this tail is related to many other physical properties of the system
through some exact relations \cite{Tan20081, Tan20082, Tan20083, tan-relation-OperatorProduct}.
The name contact comes from the fact that it is a measure of the number of pairs of fermions in two different internal states with small separations. These exact relations \cite{Tan20081, Tan20082, Tan20083, tan-relation-OperatorProduct}
have been generalized to the 1D two-component Fermi system \cite{tan-relation-1dfermi1},
the 2D two-component Fermi system \cite{Tan2005, tan-relation-2dfermi3, tan-relation-2dfermi2, tan-relation-2dfermi1}, the spin-orbit-coupled Fermi system \cite{tan-relation-coupling2, tan-relation-coupling1}, the Bose system \cite{tan-relation-bose1, tan-relation-bose2}, and the mixtures \cite{tan-relation-2dfermi3, tan-relation-bose2}.

As a zero-range interacting system, the 3D two-component Fermi gas trapped in a smooth potential has an elegant property: its energy can be expressed as a linear functional of the occupation probabilities of single-particle energy eigenstates, i.e. \cite{Tan2011}
\begin{equation}\label{eq:energyfunctional3D}
    E=\frac{\hbar^2\mathcal{I}_{3\mathrm D}}{4\pi ma}+\lim_{\epsilon_\text{M}\rightarrow\infty}\left(\sum_{\epsilon_\nu<\epsilon_\text{M}}\epsilon_\nu n_\nu-\frac{\hbar \mathcal{I}_{3\mathrm D}}{\pi^2}\sqrt{\frac{\epsilon_\text{M}}{2m}}\right),
\end{equation}
where $m$ is the mass of each fermion, $\epsilon_\nu$ is the single-particle energy of the $\nu$th single-particle level in the specified smooth external potential,
$n_\nu=n_{\nu\uparrow}+n_{\nu\downarrow}$, and $n_{\nu\uparrow}$ ($n_{\nu\downarrow}$) is the occupation probability of the spin up (down) state in the $\nu$th level.
This general functional can be regarded as a generalization of the energy of trapped non-interacting Fermi gases,
\begin{equation}
    E=\sum_{\nu\sigma}\epsilon_\nu n_{\nu\sigma}.
\end{equation}
Since the zero-range interaction model is valid for lower spatial dimensions, a straightforward idea is to generalize the energy functional \Eq{eq:energyfunctional3D} to lower dimensions. The 1D and 2D two-component Fermi gases have been studied for many years. Experimentally, one can realize them by confining the particles in some transverse directions and allowing the particles to move freely in the remaining dimensions \cite{experiment1, experiment2, experiment3}. 

In this paper, we follow the method used in Ref.~\cite{Tan2011}. We first study the one-particle density matrices of the 2D and 1D trapped two-component Fermi gases with contact interactions. We then generalize the linear energy functional \Eq{eq:energyfunctional3D} to 2D and 1D.

This paper is organized as follows. In Sec. \ref{sec2:2d-matrix}, we introduce the normalized $N$-body energy eigenstate and the 2D Bethe-Peierls boundary condition. Using the boundary condition, we expand the one-particle density matrix at small displacements. In Sec. \ref{sec3:2d-functional}, we combine the one-particle density matrix with the single-particle imaginary time propagator to find the universal energy functional in 2D:
\begin{align}\label{eq:energyfunc-2D}
    E=\lim_{\epsilon_{\mathrm{M}}\rightarrow\infty}\left(\sum_{\epsilon_\nu<\epsilon_{\mathrm{M}}}\epsilon_\nu n_\nu-\dfrac{\hbar^2\mathcal{I}_{2\mathrm{D}}}{4\pi m}\ln{\dfrac{e^{2\gamma}ma_{2\mathrm{D}}^2\epsilon_{\mathrm{M}}}{2\hbar^2}}\right),
\end{align}
where $\gamma=0.5772\cdots$ is Euler's constant, $e=2.718\cdots$ is the base of natural logarithm,
$\mathcal{I}_{2\mathrm{D}}$ is the 2D contact, $a_{2\mathrm D}$ is the 2D scattering length between two fermions
in different spin states,  $n_{\nu}=\sum_\sigma n_{\nu\sigma}$,
$\sigma=\uparrow,\downarrow$, and $n_{\nu\sigma}$ is the occupation probability of the spin-$\sigma$ state of the $\nu$th single-particle level.
If the external potential is zero, the single-particle levels reduce to plane-wave states
and \Eq{eq:energyfunc-2D} reduces to the energy theorem in Refs.~\cite{Tan2005, tan-relation-2dfermi3, tan-relation-2dfermi2, tan-relation-2dfermi1}.
One can extract the contact $\mathcal{I}_{2\mathrm{D}}$ from the asymptotic behavior of $\rho(\epsilon)$ at large $\epsilon$, where
\begin{equation}\label{eq:rho-sigma}
    \rho(\epsilon)\equiv \sum_{\nu\sigma} n_{\nu\sigma}\delta(\epsilon-\epsilon_\nu),
\end{equation}
and the coarse-grained version of $\rho(\epsilon)$ has the following asymptotic expansion at large $\epsilon$:
\beq\label{eq:rho2Dtail}
\rho(\epsilon)|_\text{cg}=\frac{\hbar^2\mathcal{I}_{2\mathrm D}}{4\pi m}\epsilon^{-2}+O(\epsilon^{-3}).
\eeq
We also derived the occupation probabilities of high energy states; see \Eq{eq:highenergystates2D}.
In Sec. \ref{sec4:1d-matrix} and Sec. \ref{sec5:1d-functional}, we do analogous calculations for the 1D two-component Fermi system and find that
\begin{eqnarray}\label{eq:energyfunc-1D}
    E &=& -\dfrac{\hbar^2a_{1\mathrm{D}}\mathcal{I}_{1\mathrm{D}}}{2m}+\sum_{\sigma\nu}\epsilon_\nu n_{\nu\sigma}, \\
    \rho(\epsilon)|_\text{cg} &=& \frac{\hbar^3\mathcal{I}_{1\mathrm{D}}}{2\sqrt{2}\pi m^{3/2}}\epsilon^{-5/2}+O(\epsilon^{-7/2}),\label{eq:rho1Dtail}
\end{eqnarray}
where $a_{1\mathrm{D}}$ is the 1D scattering length between two fermions in different spin states, and $\mathcal{I}_{1\mathrm{D}}$ is the 1D contact. 
 If the external potential is zero, the single-particle levels reduce to plane-wave states and \Eq{eq:energyfunc-1D} reduces to the energy theorem in Ref.~\cite{tan-relation-1dfermi1}.
We also derived the occupation probabilities of high energy states in 1D; see \Eq{eq:highenergystates1D}.
In Sec. \ref{sec6:summary}, 
we summarize our results and discuss the utilities and generalizations of our energy functionals.

\section{One-Particle Density Matrix in 2D}\label{sec2:2d-matrix}
\begin{figure*}
    \centering
    \subfigure[$\mathcal{D}_\eta$]{\label{fig:D_eta}
        \includegraphics[width=0.32\linewidth]{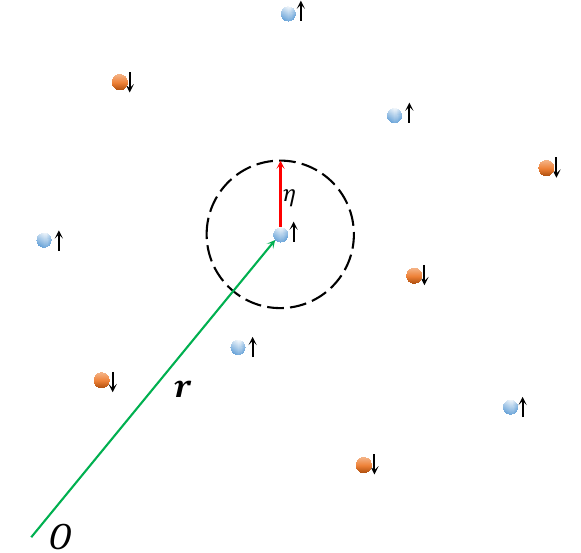}
    }
    \subfigure[The 1st subregion of $\mathcal{C}_\eta$]{\label{fig:C_eta}
        \includegraphics[width=0.32\linewidth]{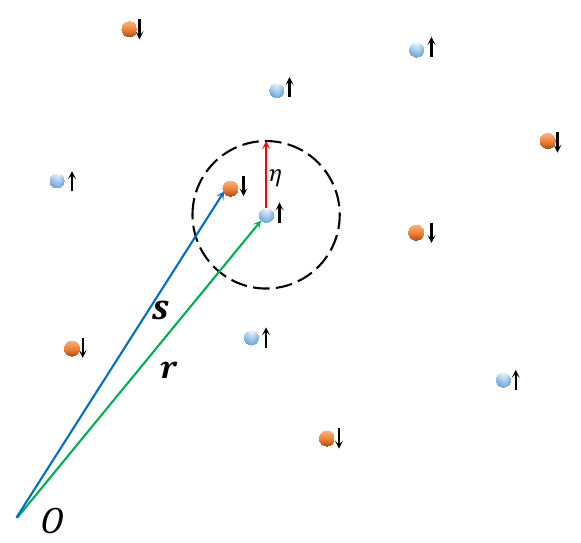}
    }
    \subfigure[2 or more spin-down fermions inside the circle of radius $\eta$ centered at $\mathbf{r}$]{\label{fig:C_more}
        \includegraphics[width=0.3\linewidth]{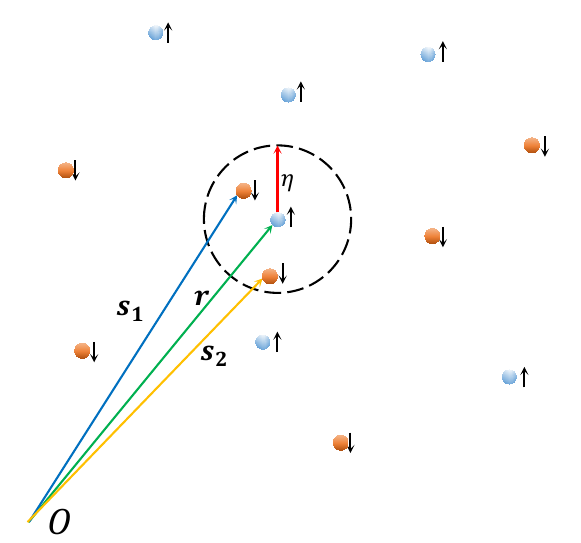}
    }
    \caption{The configuration of $N$ fermions. \ref{fig:D_eta} shows the region $\mathcal{D}_\eta$, where no spin-down fermion comes inside the circle centered at $\mathbf{r}$ with radius $\eta$. \ref{fig:C_eta} shows that a spin-down fermion at position $\mathbf{s}$ is within the circle of radius $\eta$ centered at $\mathbf{r}$. \ref{fig:C_more} shows that two or more spin-down fermions come inside the small circle, and the probability amplitudes of these situations are suppressed by Fermi statistics.}
\end{figure*}

We consider a trapped two-component Fermi system in 2D, with $N_\uparrow$ spin-up fermions and $N_\downarrow$ spin-down fermions.
The total number is $N=N_\downarrow+N_\uparrow$. Here the trapping potential $V(\vect r)$ is assumed to be smooth. First we calculate the one-particle density matrix. Consider a normalized $N$-body energy eigenstate
\begin{eqnarray}\label{eq:phi2D}
    \ket{\Phi} &=&(N_\uparrow!N_\downarrow!)^{-1/2}\int D_1^\uparrow D_1^\downarrow\Phi(\mathbf{r}_1\dots\mathbf{r}_{N_\uparrow}\mathbf{s}_1\dots\mathbf{s}_{N_\downarrow}) \notag \\
    & &\times\psi_\uparrow^\dagger(\mathbf{r}_1)\dots\psi_\uparrow^\dagger(\mathbf{r}_{N_\uparrow})\psi_\downarrow^\dagger(\mathbf{s}_1)\dots\psi_\downarrow^\dagger(\mathbf{s}_{N_\downarrow})\ket{0},
\end{eqnarray}
where $\vect r_1,\dots,\vect r_{N_\uparrow}$ are the position vectors of the spin-up fermions,
$\vect s_1,\dots,\vect s_{N_\downarrow}$ are the position vectors of the spin-down fermions,
$\psi_\uparrow^\dagger(\vect r)$ is the creation operator for a spin-up fermion at position $\vect r$,
$\psi_\downarrow^\dagger(\vect s)$ is the creation operator for a spin-down fermion at position $\vect s$,
 $D_i^\uparrow\equiv\prod_{\mu=i}^{N_\uparrow}\mathrm{d}^2r_\mu$, $D_i^\downarrow\equiv\prod_{\mu=i}^{N_\downarrow}\mathrm{d}^2s_\mu$,
 and $\Phi(\vect r_1\dots\vect r_{N_\uparrow}\vect s_1\dots\vect s_{N_\downarrow})$ is the $N$-body wave function
 which is antisymmetric under the interchange of the positions of any two spin-up (spin-down) fermions.
When $\mathbf{r}_1$ and $\mathbf{s}_1$ are close, the wave function 
satisfies the 2D Bethe-Peierls boundary condition
\begin{eqnarray}\label{eq:B-P BC}
    \Phi &=& A\left(\frac{\mathbf{r}_1+\mathbf{s}_1}{2};\mathbf{r}_2\dots\mathbf{r}_{N_\uparrow}\mathbf{s}_2\dots\mathbf{s}_{N_\downarrow}\right)\nonumber \\
    & &\times \frac{1}{2\pi}\ln{\frac{a_{2\mathrm{D}}}{|\mathbf{r}_1-\mathbf{s}_1|}+O(|\mathbf{r}_1-\mathbf{s}_1|)},
\end{eqnarray}
where $a_{2\mathrm{D}}$ is the two-dimensional $s$-wave scattering length,
and $A$ is a function of $(N-1)$ position vectors. The one-particle density matrix for the spin-$\sigma$ fermions is defined as
\beq\label{eq:one-particle-density-matrix}
p_\sigma(\vect r,\vect r+\vect b)=\bra{\Phi}\psi_\sigma^\dagger(\vect r)\psi_\sigma(\vect r+\vect b)\ket{\Phi}.
\eeq
In particular, by substituting \Eq{eq:phi2D} into the above definition, we find that
\begin{eqnarray}
    p_\uparrow(\mathbf{r},\mathbf{r}+\mathbf{b}) &=& N_\uparrow\int D_2^\uparrow D_1^\downarrow \Phi^*(\mathbf{r},\mathbf{r}_2\dots\mathbf{r}_{N_\uparrow}\mathbf{s}_1\dots\mathbf{s}_{N_\downarrow})\notag \\
    & & \times\Phi(\mathbf{r+b},\mathbf{r}_2\dots\mathbf{r}_{N_\uparrow}\mathbf{s}_1\dots\mathbf{s}_{N_\downarrow}).
\end{eqnarray}
We will expand $p_\uparrow(\mathbf{r},\mathbf{r}+\mathbf{b})$ through order ${O(b^3)}$ at small distance $b$. Since $\Phi$ is singular when two fermions in different spin states are close, we divide the $2(N-1)$-dimensional integration domain into two regions: $\mathcal{C}_\eta$ and $\mathcal{D}_\eta$. $\mathcal{D}_\eta$ is the region in which every spin-down fermion lies outside of the circle of radius $\eta$ centered at $\mathbf{r}$, that is, $|\mathbf{s}_\mu-\mathbf{r}|>\eta$ for $\mu=1, \dots, N_\downarrow$, which is shown in \Fig{fig:D_eta}. $\mathcal{C}_\eta$ is the complement of $\mathcal{D}_\eta$. We set $\eta$ small but $\eta>b$. 
In $\mathcal{C}_\eta$ the cases that two or more spin-down fermions come inside the small circle of radius $\eta$ centered at $\mathbf{r}$ are possible, but the contributions from these cases are suppressed by Fermi statistics and are of higher order than $b^4$, see \Fig{fig:C_more}. Next, we calculate the integrals in these two regions and add them up, then the dependencies on $\eta$ will be canceled.

In $\mathcal{D}_\eta$, we expand $\Phi$ in powers of $\mathbf{b}$ as
\begin{eqnarray}
    \Phi(\mathbf{r+b},\mathbf{R}) &=& \Phi(\mathbf{r},\mathbf{R})+\nabla_\mathbf{r}\Phi(\mathbf{r},\mathbf{R})\cdot\mathbf{b}\nonumber \\
    & & +\dfrac{1}{2}\sum_{i,j=1}^2\dfrac{\partial^2}{\partial r_i\partial r_j}\Phi(\mathbf{r},\mathbf{R})b_i b_j\nonumber \\
    & &+T_{\mathbf{b}}+O(b^4),
\end{eqnarray}
where $\mathbf{R}\equiv (\mathbf{r}_2\dots\mathbf{r}_{N_\uparrow}\mathbf{s}_1\dots\mathbf{s}_{N_\downarrow})$ and $T_{\mathbf{b}}\equiv\dfrac{1}{3!}\sum_{i,j,k=1}^2\frac{\partial^3}{\partial r_i\partial r_j\partial r_k}\Phi(\mathbf{r},\mathbf{R})b_i b_j b_k$.
Let $I_\mathcal{D}$ be the integral evaluated in $\mathcal{D}_\eta$, and $I_\mathcal{C}$ be the integral evaluated in $\mathcal{C}_\eta$. We find
\begin{eqnarray}
    I_\mathcal{D} &=& n_\uparrow^\mathcal{D}(\mathbf{r})+\mathbf{u}_\uparrow(\mathbf{r})\cdot\mathbf{b}+\dfrac{1}{2}\sum_{i,j=1}^2v_{\uparrow,ij}(\mathbf{r})b_i b_j \nonumber \\
    & & +T'_{\mathbf{b}}+O(b^4),
\end{eqnarray}
where
\begin{align}
    n_\uparrow^\mathcal{D}(\mathbf{r}) &= N_\uparrow\lim_{\eta\to 0}
    \int_{\mathcal{D}_\eta}D_2^\uparrow D_1^\downarrow |\Phi(\mathbf{r},\mathbf{R})|^2, \\
    \mathbf{u}_\uparrow(\mathbf{r}) &= N_\uparrow\lim_{\eta\to 0}
    \int_{\mathcal{D}_\eta}D_2^\uparrow D_1^\downarrow\Phi^*(\mathbf{r},\mathbf{R})\nabla_\mathbf{r}\Phi(\mathbf{r},\mathbf{R}), \\
    v_{\uparrow,ij}(\mathbf{r}) &= N_\uparrow\lim_{\eta\to 0}
    \int_{\mathcal{D}_\eta}D_2^\uparrow D_1^\downarrow\Phi^*(\mathbf{r},\mathbf{R})\dfrac{\partial^2}{\partial r_i\partial r_j}\Phi(\mathbf{r},\mathbf{R}), \\
    T'_{\mathbf{b}} &= N_\uparrow\lim_{\eta\to 0}
    \int_{\mathcal{D}_\eta}D_2^\uparrow D_1^\downarrow\Phi^*(\mathbf{r},\mathbf{R})T_{\mathbf{b}}.
\end{align}
To calculate the contributions from the region $\mathcal{C}_\eta$, we use the Bethe-Peierls boundary condition (\ref{eq:B-P BC}). The region $\mathcal{C}_\eta$ can be approximately partitioned into $N_\downarrow$ subregions,
and in the $\mu$th subregion ($\mu=1,\dots,N_\downarrow$) $\vect s_\mu$ is within the circle of radius $\eta$ centered at $\vect r$.
The contributions to $I_\mathcal{C}$ from these subregions are equal due to Fermi statistics.
In the first subregion (shown in \Fig{fig:C_eta}) we have
\begin{eqnarray}
    \Phi(\mathbf{r},\mathbf{R}') &=& A\left(\frac{\mathbf{r}+\mathbf{s}}{2};\mathbf{R}'\right)\times\dfrac{1}{2\pi}\ln{\dfrac{a_{2\mathrm{D}}}{|\mathbf{r}-\mathbf{s}|}}\nonumber \\
    & & +O(|\mathbf{r}-\mathbf{s}|), \\
    \Phi(\mathbf{r}+\mathbf{b},\mathbf{R}') &=& A\left(\frac{\mathbf{r}+\mathbf{s}+\mathbf{b}}{2};\mathbf{R}'\right)\times\dfrac{1}{2\pi}\ln{\dfrac{a_{2\mathrm{D}}}{|\mathbf{r}+\mathbf{b}-\mathbf{s}|}}\nonumber \\
    & &+O(|\mathbf{r}+\mathbf{b}-\mathbf{s}|),
\end{eqnarray}
where $\mathbf{s}\equiv \mathbf{s}_1$ and $\mathbf{R}'\equiv (\mathbf{r}_2\dots\mathbf{r}_{N_\uparrow}\mathbf{s}_2\dots\mathbf{s}_{N_\downarrow})$. We then do the following expansions:
\begin{equation}
    A\left(\frac{\mathbf{r}+\mathbf{s}}{2};\mathbf{R}'\right) = A(\mathbf{r};\mathbf{R}')-\dfrac{\mathbf{q}}{2}\cdot\nabla_\vect r A+O(q^2), 
\end{equation}
\begin{align}
    A\left(\frac{\mathbf{r}+\mathbf{s}+\mathbf{b}}{2};\mathbf{R}'\right)=\, & A(\mathbf{r}; \mathbf{R}')+\left(\dfrac{\mathbf{b}}{2}-\dfrac{\mathbf{q}}{2}\right)\cdot\nabla_\vect r A\nonumber \\
    & +O(|\mathbf{b}-\mathbf{q}|^2),
\end{align}
where $\mathbf{q}=\mathbf{r}-\mathbf{s}$. So we have
\begin{equation}
    I_\mathcal{C}
    = N_\uparrow N_\downarrow
    \int D_2^\uparrow D_2^\downarrow\int_{q<\eta}\mathrm{d}^2q F_\mathbf{b}+O(b^4),
\end{equation}
where
\begin{eqnarray}
    F_\mathbf{b} &=& \dfrac{1}{4\pi^2}\ln{\dfrac{a_{2\mathrm{D}}}{q}}\ln{\dfrac{a_{2\mathrm{D}}}{|\mathbf{q}+
    \mathbf{b}|}}\left(A^*-\nabla_\mathbf{r}A^*\cdot\dfrac{\mathbf{q}}{2}\right)\nonumber \\
    & & \times\left[A+\nabla_\mathbf{r}A\cdot\left(\dfrac{\mathbf{b}}{2}-\dfrac{\mathbf{q}}{2}\right)\right].
\end{eqnarray}
Carrying out the integral $I_\mathcal{C}$ and adding it to $I_\mathcal{D}$, we get 
\begin{align}
    p_\uparrow(\mathbf{r},\mathbf{r}+\mathbf{b}) =& \, I_\mathcal{C}+I_\mathcal{D} \notag \\
    =& \, n_\uparrow(\mathbf{r})+\mathbf{u}_\uparrow(\mathbf{r})\cdot\mathbf{b}+\frac{1}{8\pi}C_{2\mathrm{D}}(\mathbf{r})b^2\ln{\frac{b}{a_{2\mathrm{D}}e}}\nonumber \\
    & +\dfrac{1}{2}\sum_{i,j=1}^2v_{\uparrow,ij}(\mathbf{r})b_i b_j \nonumber \\
    & +\dfrac{1}{16\pi}b^2\left(\dfrac{1}{2}\ln{\dfrac{b}{a_{2\mathrm{D}}}}-\dfrac{3}{8}\right)\mathbf{w}^*\cdot\mathbf{b} \nonumber \\
    & +\dfrac{1}{16\pi}b^2\left(\dfrac{3}{2}\ln{\dfrac{b}{a_{2\mathrm{D}}}}-\dfrac{11}{8}\right)\mathbf{w}\cdot\mathbf{b}\nonumber \\
    & +T'_{\mathbf{b}}+O(b^4),
\end{align}
where
\begin{eqnarray}
    n_\uparrow(\mathbf{r}) 
    &=& N_\uparrow\int D_2^\uparrow D_1^\downarrow |\Phi(\mathbf{r},\mathbf{R})|^2, \\
    C_{2\mathrm{D}}(\mathbf{r}) &\equiv& N_\uparrow N_\downarrow\int D_2^\uparrow D_2^\downarrow\ |A(\mathbf{r};\mathbf{R}')|^2, \\
    \mathbf{w}(\mathbf{r}) &\equiv& N_\uparrow N_\downarrow\int D_2^\uparrow D_2^\downarrow\ A^*(\mathbf{r};\mathbf{R}')\nabla_\mathbf{r}A(\mathbf{r};\mathbf{R}').
\end{eqnarray}
$n_\uparrow(\mathbf{r})$ is the spatial density of spin-up fermions at $\mathbf{r}$, $C_{2\mathrm{D}}(\mathbf{r})$ is the 2D contact density, 
and $\mathbf{w}(\mathbf{r})$ is related to the center-of-mass motion of small-distance pairs of fermions in different spin states.
We can also find a similar expansion for $p_\downarrow(\mathbf{r},\mathbf{r}+\mathbf{b})$.

\section{Universal Energy Functional in 2D}\label{sec3:2d-functional}
We define an absolutely convergent series
\begin{equation}
    J_\sigma(\beta)\equiv\sum_\nu n_{\nu\sigma}e^{-\beta \epsilon_\nu}=\sum_\nu\expval{c^\dagger_{\nu\sigma}c_{\nu\sigma}}{\Phi}e^{-\beta \epsilon_\nu},
\end{equation}
where $\beta$ satisfies Re$\beta\ge0$, $\ket{\Phi}$ is an $N$-body energy eigenstate,
\beq\label{eq:nnusigma}
n_{\nu\sigma}=\bra{\Phi}c_{\nu\sigma}^\dagger c_{\nu\sigma}\ket{\Phi}
\eeq
is the occupation probability of the spin-$\sigma$ state of the $\nu$th single-particle level,
\begin{equation}\label{eq:fermion-annihilation}
    c_{\nu\sigma}=\int\mathrm{d}^2r\phi^*_\nu(\mathbf{r})\psi_\sigma(\mathbf{r})
\end{equation}
is the fermion annihilation operator of such a single-particle state, and $\phi_\nu(\mathbf{r})$ is the wave function of the $\nu$th single-particle level in the trapping potential $V(\vect r)$ and satisfies the single-particle Schr\"{o}dinger equation
\begin{equation}\label{eq:single-particleSchrodinger2D}
    \left[-\frac{\hbar^2}{2m}\nabla^2+V(\mathbf{r})\right]\phi_\nu(\mathbf{r})=\epsilon_\nu\phi_\nu(\mathbf{r})
\end{equation}
and the normalization condition
\beq
\int |\phi_\nu(\vect r)|^2\mathrm{d}^2r=1.
\eeq
We rewrite $J_\sigma(\beta)$ as
\begin{equation}\label{eq:seriesdef}
    J_\sigma(\beta)=\int\mathrm{d}^2r\mathrm{d}^2r'U_\beta(\mathbf{r},\mathbf{r}')p_\sigma(\mathbf{r},\mathbf{r}'),
\end{equation}
where $U_\beta(\mathbf{r},\mathbf{r}')=\sum_\nu e^{-\beta\epsilon_\nu}\phi_\nu(\mathbf{r})\phi^*_\nu(\mathbf{r}')$ is the propagator of a single particle moving in the potential $V(\mathbf{r})$ whthin a time $-i\hbar\beta$. For a small positive $\beta$, at $|\mathbf{r}-\mathbf{r}'|\gg\hbar\sqrt{\beta/m}$ the propagator is exponentially suppressed, while at $|\mathbf{r}-\mathbf{r}'|\sim\hbar\sqrt{\beta/m}$ we have a short imaginary-time expansion
\begin{eqnarray}
    U_\beta(\mathbf{r},\mathbf{r}') &=& \dfrac{m}{2\pi\hbar^2\beta}\left[1-\frac{V(\mathbf{r})+V(\vect r')}{2}\beta\right]\nonumber \\
    & & \times\exp\left[-\dfrac{m(\mathbf{r}-\mathbf{r}')^2}{2\hbar^2\beta}\right]+O(\beta).
\end{eqnarray}
Recall that when $|\mathbf{r}-\mathbf{r}'|$ is small, we also have an expansion of $p_\sigma(\mathbf{r},\mathbf{r}')$. Defining $G_\sigma=U_\beta(\vect r,\vect r')p_\sigma(\vect r,\vect r')$, we have
\begin{eqnarray}
    G_\sigma &=& \dfrac{m}{2\pi\hbar^2\beta}\left[1-\frac{V(\mathbf{r})+V(\vect r')}{2}\beta\right]\exp\left[-\dfrac{mb^2}{2\hbar^2\beta}\right] \nonumber \\
    & & \times\Bigg[n_\uparrow(\mathbf{r})+\mathbf{u}_\uparrow(\mathbf{r})\cdot\mathbf{b}+\sum_{i,j=1}^2v_{\uparrow,ij}(\mathbf{r})\dfrac{b_i b_j}{2}\nonumber \\
    & & +\dfrac{b^2C_{2\mathrm{D}}(\mathbf{r})}{8\pi}\ln{\dfrac{b}{a_{2\mathrm{D}}e}}+\dfrac{b^2\mathbf{w}\cdot\mathbf{b}}{16\pi}\left(\dfrac{3}{2}\ln{\dfrac{b}{a_{2\mathrm{D}}}}-\dfrac{11}{8}\right) \nonumber \\
    & & +\dfrac{b^2\mathbf{w}^*\cdot\mathbf{b}}{16\pi}\left(\dfrac{1}{2}\ln{\dfrac{b}{a_{2\mathrm{D}}}}-\dfrac{3}{8}\right)\Bigg]+O(\beta),
\end{eqnarray}
where $\vect b=\mathbf{r}'-\mathbf{r}$. Substituting the above result into \Eq{eq:seriesdef}, we find
\begin{eqnarray}\label{eq:seriesexpr}
    J_\sigma(\beta) &=& N_\sigma-\beta\int\mathrm{d}^2rV(\mathbf{r})n_\sigma(\mathrm{r})\nonumber \\
    & & -\dfrac{\hbar^2\mathcal{I}_{2\mathrm{D}}\beta}{8\pi m}\left(1+\gamma+\ln{\dfrac{ma_{2\mathrm{D}}^2}{2\hbar^2\beta}}\right)\nonumber \\
    & & +\dfrac{\hbar^2\beta}{2m}\int\mathrm{d}^2r\sum_{i=1}^2v_{\sigma,ii}(\mathbf{r})+O(\beta^2), 
\end{eqnarray}
where
\begin{eqnarray}
    N_\sigma &=& \int\mathrm{d}^2rn_\sigma(\mathbf{r}), \\
    \mathcal{I}_{2\mathrm{D}} &=& 
    \int\mathrm{d}^2rC_{2\mathrm{D}}(\mathbf{r}).
\end{eqnarray}
Outside of the tiny range of two-body interactions, the $N$-body Schr\"{o}dinger equation is simplified as
\begin{eqnarray}\label{eq:Sch_eq}
    E\Phi &=& \sum_{\mu=1}^{N_\uparrow}\left[-\dfrac{\hbar^2}{2m}\nabla^2_{\mathbf{r}_\mu}+V(\mathbf{r}_\mu)\right]\Phi \nonumber \\
    & &+\sum_{\mu'=1}^{N_\downarrow}\left[-\dfrac{\hbar^2}{2m}\nabla^2_{\mathbf{s}_{\mu'}}+V(\mathbf{s}_{\mu'})\right]\Phi, 
\end{eqnarray}
where $\mathbf{r}_\mu\ne\mathbf{s}_{\mu'}$ for all $\mu, \mu'$. Multiplying both sides of \Eq{eq:Sch_eq} by $\Phi^*$, integrating them 
over $\vect r_1,\dots,\vect r_{N_\uparrow},\vect s_1,\dots,\vect s_{N_\downarrow}$ for $\mathbf{r}_\mu\ne\mathbf{s}_{\mu'}$ for all $\mu, \mu'$, we get
\begin{equation}
    \sum_\sigma\int\mathrm{d}^2r\left[V(\mathbf{r})n_\sigma(\mathbf{r})-\dfrac{\hbar^2}{2m}\sum_{i=1}^2v_{\sigma,ii}(\mathbf{r})\right]=E.
\end{equation}
Summing \Eq{eq:seriesexpr} over $\sigma$, we find
\begin{align}
    \sum_\sigma J_\sigma(\beta) =&\, \sum_{\nu\sigma}n_{\nu\sigma}e^{-\beta\epsilon_\nu} \nonumber \\
    =&\, N-\dfrac{\hbar^2\mathcal{I}_{2\mathrm{D}}\beta}{8\pi m}\left(1+\gamma+\ln{\dfrac{ma_{2\mathrm{D}}^2}{2\hbar^2\beta}}\right) \nonumber \\
    & -\beta\sum_\sigma\int\mathrm{d}^2r\left[V(\mathbf{r})n_\sigma-\dfrac{\hbar^2}{2m}\sum_{i=1}^2v_{\sigma,ii}\right] \nonumber \\
    &+O(\beta^2)\nonumber\\
    =&\, N-E\beta-\dfrac{\hbar^2\mathcal{I}_{2\mathrm{D}}\beta}{4\pi m}\left(1+\gamma+\ln{\dfrac{ma_{2\mathrm{D}}^2}{2\hbar^2\beta}}\right)\nonumber \\
    &+O(\beta^2).\label{eq:Jsigmaexpand}
\end{align}
Let
\begin{equation}
    \rho(\epsilon)\equiv\sum_{\sigma}\rho_{\sigma}(\epsilon)=\sum_{\nu\sigma}n_{\nu\sigma}\delta(\epsilon-\epsilon_\nu).
\end{equation}
Equation \eqref{eq:Jsigmaexpand} can be rewritten as
\begin{align}\label{eq:rho2Dtransform}
\int_{-\infty}^{+\infty}\rho(\epsilon)e^{-\beta\epsilon}\mathrm{d}\epsilon= &\,N-\dfrac{\hbar^2\mathcal{I}_{2\mathrm{D}}\beta}{4\pi m}\left(1+\gamma+\ln{\dfrac{ma_{2\mathrm{D}}^2}{2\hbar^2\beta}}\right)\nonumber \\
    &-E\beta+O(\beta^2).
\end{align}
Setting $\beta=\eta+i s$ where $\eta$ is a positive infinitesimal and $s$ is real, we see that the above equation
shows the Fourier transform of $\rho(\epsilon)$ at small $s$, and this Fourier transform has a singular term proportional to $s \ln s$.
This singular term is caused by a power law tail of the coarse-grained version of $\rho(\epsilon)$ at $\epsilon\to\infty$.
Taking the inverse Fourier transform of this singular term, we find the power law tail shown in \Eq{eq:rho2Dtail}. 

Applying $\frac{\mathrm{d}}{\mathrm{d}\beta}$ to both sides of \Eq{eq:rho2Dtransform}, we find
\begin{equation}
    E 
    =\int_{-\infty}^\infty\rho(\epsilon)\epsilon e^{-\beta\epsilon}\mathrm{d}\epsilon-\dfrac{\hbar^2\mathcal{I}_{2\mathrm{D}}}{4\pi m}\left(\gamma+\ln{\dfrac{a_{2\mathrm{D}}^2m}{2\hbar^2\beta}}\right)+O(\beta).
\end{equation}
We divide the domain of integration over $\epsilon$ into two regions: one is $(-\infty,\epsilon_\mathrm{M})$ and the other is $(\epsilon_\mathrm{M},\infty)$, where $\epsilon_\mathrm{M}$ is an energy scale such that $\epsilon_\mathrm{M}$ is very large but $\epsilon_\mathrm{M}\beta\ll1$. In $(-\infty,\epsilon_\mathrm{M})$ we have
\beq
\int_{-\infty}^{\epsilon_\mathrm{M}}\rho(\epsilon)\epsilon e^{-\beta\epsilon}\mathrm{d}\epsilon\approx\int_{-\infty}^{\epsilon_\mathrm{M}}\rho(\epsilon)\epsilon \mathrm{d}\epsilon,
\eeq
while in $(\epsilon_\mathrm{M},\infty)$ we use \Eq{eq:rho2Dtail} to do the integral:
\begin{align}
    \int_{\epsilon_\mathrm{M}}^\infty\rho(\epsilon)\epsilon e^{-\beta\epsilon}\mathrm{d}\epsilon &=\dfrac{\hbar^2\mathcal{I}_{2\mathrm{D}}}{4\pi m}\int_{\epsilon_\mathrm{M}}^\infty\epsilon^{-1}e^{-\beta\epsilon}\mathrm{d}\epsilon \notag \\
    &= \dfrac{\hbar^2\mathcal{I}_{2\mathrm{D}}}{4\pi m}\mathrm{\Gamma}(0,\epsilon_{\mathrm{M}}\beta) \nonumber\\
    &= \dfrac{\hbar^2\mathcal{I}_{2\mathrm{D}}}{4\pi m}[-\gamma-\ln{(\epsilon_{\mathrm{M}}\beta)}]+O(\epsilon_{\mathrm{M}}\beta).
\end{align}
Thus, taking $\beta\rightarrow0$, we get
\begin{align}
    E &= \lim_{\epsilon_{\mathrm{M}}\rightarrow\infty}\left[\int_{-\infty}^{\epsilon_{\mathrm{M}}}\rho(\epsilon)\epsilon \mathrm{d}\epsilon-\dfrac{\hbar^2\mathcal{I}_{2\mathrm{D}}}{4\pi m}\ln{\dfrac{e^{2\gamma}ma_{2\mathrm{D}}^2\epsilon_{\mathrm{M}}}{2\hbar^2}}\right] \notag \\
    &=\lim_{\epsilon_{\mathrm{M}}\rightarrow\infty}\left(\sum_{\epsilon_\nu<\epsilon_{\mathrm{M}}}\epsilon_\nu n_\nu-\dfrac{\hbar^2\mathcal{I}_{2\mathrm{D}}}{4\pi m}\ln{\dfrac{e^{2\gamma}ma_{2\mathrm{D}}^2\epsilon_{\mathrm{M}}}{2\hbar^2}}\right),
\end{align}
which is \Eq{eq:energyfunc-2D}.


According to Eqs. \eqref{eq:one-particle-density-matrix}, \eqref{eq:nnusigma}, and \eqref{eq:fermion-annihilation}, we have
\begin{equation}\label{eq:nnusigmaintegral}
    n_{\nu\sigma}=\int \text{d}^2r\int\text{d}^2b\, \phi_\nu(\mathbf{r})\phi_\nu^*(\mathbf{r}+\mathbf{b})p_\sigma(\mathbf{r},\mathbf{r}+\mathbf{b}).
\end{equation}
When $\epsilon_\nu$ is large, the integrand as a function of $\mathbf{b}$ oscillates rapidly, which implies that the only important contribution is from the singular term in the expansion of $p_\sigma(\vect r,\vect r+\vect b)$ at $\mathbf{b}\rightarrow0$ \cite{Tan2011},
and this singular term is $\frac{1}{8\pi}C_{2\mathrm{D}}(\mathbf{r})b^2\ln{\frac{b}{a_{2\mathrm{D}}e}}$. Since $\phi_\nu(\mathbf{r})$ satisfies the single-particle Schr\"{o}dinger equation, \Eq{eq:single-particleSchrodinger2D},
we have
\beq\label{eq:phinu*r+b}
\phi^*_\nu(\mathbf{r}+\mathbf{b})\approx(\hbar^2/2m\epsilon_\nu)^2\nabla^4_\mathbf{b}\phi^*_\nu(\mathbf{r}+\mathbf{b})
\eeq
with relative error $\sim O(\epsilon_\nu^{-1})$ at $b\sim\sqrt{\hbar^2/2m\epsilon_\nu}$. Substituting \Eq{eq:phinu*r+b}
into \Eq{eq:nnusigmaintegral} and carrying out the integral over $\vect b$, we find
\begin{equation}\label{eq:highenergystates2D}
    n_{\nu\sigma}=\frac{1}{k_\nu^4}\int\text{d}^2rC_{2\mathrm{D}}(\mathbf{r})|\phi_\nu(\mathbf{r})|^2+O(\epsilon_\nu^{-5/2}),
\end{equation}
where $k_\nu=\sqrt{2m\epsilon_\nu}/\hbar$.

\section{One-Particle Density Matrix in 1D}\label{sec4:1d-matrix}
The calculation procedure in 1D is similar to the one in 2D. We define a normalized $N$-body energy eigenstate $\ket{\Phi}$ in 1D,
\begin{align}\label{eq:Nwavefunction1D}
    \ket{\Phi} =&(N_\uparrow!N_\downarrow!)^{-1/2}\int \Tilde{D}_1^\uparrow \Tilde{D}_1^\downarrow\Phi(x_1\dots x_{N_\uparrow}y_1\dots y_{N_\downarrow}) \notag \\
    &\times\psi_\uparrow^\dagger(x_1)\dots\psi_\uparrow^\dagger(x_{N_\uparrow})\psi_\downarrow^\dagger(y_1)\dots\psi_\downarrow^\dagger(y_{N_\downarrow})\ket{0}.
\end{align}
where $N=N_{\uparrow}+N_\downarrow$, $N_\uparrow$ is the number of spin-up fermions, $N_\downarrow$ is the number of spin-down fermions,
$x_1,\dots,x_{N_\uparrow}$ are the coordinates of the spin-up fermions,
$y_1,\dots,y_{N_\downarrow}$ are the coordinates of the spin-down fermions,
$\psi_\uparrow^\dagger(x)$ is the creation operator for a spin-up fermion at position $x$,
$\psi_\downarrow^\dagger(y)$ is the creation operator for a spin-down fermion at position $y$,
$\Tilde{D}_i^{\uparrow}\equiv\prod_{\mu=i}^{N_\uparrow}\mathrm{d}x_\mu$, $\Tilde{D}_i^\downarrow\equiv\prod_{\mu=i}^{N_\downarrow}\mathrm{d}y_\mu$, and $\Phi(x_1\dots x_{N_\uparrow}y_1\dots y_{N_\downarrow})$ is the $N$-body wave function which is antisymmetric under the interchange of any two spin-up (spin-down) fermions. The 1D Bethe-Peierls boundary condition is
\begin{eqnarray}\label{eq:B-P BC1D}
    \ket{\Phi}&=&A\left(\frac{x_1+y_1}{2};x_2\dots x_{N_\uparrow}y_2\dots y_{N_\downarrow}\right)\nonumber \\
    & & \times \left(1-\frac{|x_1-y_1|}{a_{1\mathrm{D}}}\right)+O(|x_1-y_1|^2),
\end{eqnarray}
which is satisfied by the wave function when $x_1$ and $y_1$ are close.
The one-particle density matrix for spin-$\sigma$ fermions in 1D is defined as
\begin{equation}
    p_\sigma(x,x+b) = \bra{\Phi}\psi_\sigma^\dagger(x)\psi_\sigma(x+b)\ket{\Phi}.
\end{equation}
For spin-up fermions, we substitute \Eq{eq:Nwavefunction1D} into the above definition and find
\begin{align}
    p_\uparrow(x,x+b)=&\, N_\uparrow\int \Tilde{D}_2^\uparrow \Tilde{D}_1^\downarrow \Phi^*(x, x_2\dots x_{N_\uparrow}y_1\dots y_{N_\downarrow})\nonumber \\
    &\times\Phi(x+b, x_2\dots x_{N_\uparrow}y_1\dots y_{N_\downarrow}).
\end{align}
After finishing calculations analogous to those for the 2D one-particle density matrix, we find 
\begin{align}
     p_\uparrow(x,x+b) =& n_\uparrow(x)+u_\uparrow(x)b+\frac{1}{2}v_{\uparrow}(x)b^2\nonumber \\
     & +C_{1\mathrm D}(x)\left(-\frac{b^2a_{1\mathrm{D}}}{4}+\frac{|b|^3}{12}\right) \nonumber \\
     & +w(x)\frac{2b^3}{3a_{1\mathrm{D}}}+w^*(x)\frac{b^3}{6a_{1\mathrm{D}}}+T'_{b}+O(b^4),
\end{align}
where 
\begin{align}
    n_\uparrow(x) &= N_\uparrow\int \Tilde{D}_2^\uparrow \Tilde{D}_1^\downarrow |\Phi(x,\mathbf{X})|^2, \\
    u_\uparrow(x) &= N_\uparrow\lim_{\eta\to 0}
    \int_{\mathcal{D}_\eta}\Tilde{D}_2^\uparrow \Tilde{D}_1^\downarrow\Phi^*(x,\mathbf{X})\frac{\partial}{\partial x}\Phi(x,\mathbf{X}), \\
    v_{\uparrow}(x) &= N_\uparrow\lim_{\eta\to 0}
    \int_{\mathcal{D}_\eta}\Tilde{D}_2^\uparrow \Tilde{D}_1^\downarrow\Phi^*(x, \mathbf{X})\dfrac{\partial^2}{\partial x^2}\Phi(x,\mathbf{X}), \\
    T'_{b} &= \frac{N_\uparrow b^3}{3!}\lim_{\eta\to 0}
    \int_{\mathcal{D}_\eta}\Tilde{D}_2^\uparrow \Tilde{D}_1^\downarrow\Phi^*(x, \mathbf{X})\dfrac{\partial^3}{\partial x^3}\Phi(x,\mathbf{X}), \\
    C_{1\mathrm D}(x) &\equiv \frac{4N_\uparrow N_\downarrow}{a_{1\mathrm D}^2}\int\Tilde{D}_2^\uparrow \Tilde{D}_2^\downarrow\ |A(x;\mathbf{X}')|^2, \\
    w(x) &\equiv N_\uparrow N_\downarrow\int\Tilde{D}_2^\uparrow \Tilde{D}_2^\downarrow\ A^*(x; \mathbf{X}')\frac{\partial A(x; \mathbf{X}')}{\partial x},
\end{align}
and $\mathbf{X}\equiv (x_2\dots x_{N_\uparrow}y_1\dots y_{N_\downarrow})$, $\mathbf{X}'\equiv (x_2\dots x_{N_\uparrow}y_2\dots y_{N_\downarrow})$.
$n_\uparrow(x)$ is the spatial density of spin-up fermions at position $x$, $C_{1\mathrm{D}}(x)$ is the 1D contact density, and $w(x)$ is related to the center-of-mass motion of small-distance pairs of fermions in different spin states.
We can also find a similar expansion for $p_\downarrow(x,x+b)$.

\section{Universal Energy Functional in 1D}\label{sec5:1d-functional}
We define $J_\sigma(\beta)$ in 1D:
\begin{equation}
    J_\sigma(\beta)\equiv\sum_\nu n_{\nu\sigma}e^{-\beta \epsilon_\nu}=\sum_\nu\expval{c^\dagger_{\nu\sigma}c_{\nu\sigma}}{\Phi}e^{-\beta \epsilon_\nu},
\end{equation}
where
\begin{equation}
    c_{\nu\sigma}=\int_{-\infty}^\infty\phi^*_\nu(x)\psi_\sigma(x)\mathrm{d}x,
\end{equation}
and $\phi_\nu(x)$ is the wave function of the $\nu$th single-particle  level  in  the  trapping  potential $V(x)$  and satisfies 
the single-particle Schr\"{o}dinger equation
\beq
\left[-\frac{\hbar^2}{2m}\frac{\mathrm{d}^2}{\mathrm{d}x^2}+V(x)\right]\phi_\nu(x)=\epsilon_\nu\phi_\nu(x)
\eeq
and the normalization condition
\beq
\int_{-\infty}^\infty|\phi_\nu(x)|^2\mathrm{d} x=1.
\eeq

We can rewrite $J_\sigma(\beta)$ as
\begin{equation}\label{eq:seriesdef1}
    J_\sigma(\beta)=\int_{-\infty}^\infty\mathrm{d}x\int_{-\infty}^\infty\mathrm{d}x'U_\beta(x,x')p_\sigma(x,x'),
\end{equation}
where $U_\beta(x,x')=\sum_\nu e^{-\beta\epsilon_\nu}\phi_\nu(x)\phi^*_\nu(x')$ is the propagator of a single particle moving in the potential $V(x)$ within a time $-i\hbar\beta$. 
We find 
\begin{align}\label{eq:seriesexpr1D}
    J_\sigma(\beta) =& N_\sigma-\beta\int\mathrm{d}x V(x)n_\sigma(x)-\dfrac{\hbar^2a_{1\mathrm D}\mathcal{I}_{1\mathrm D}\beta}{4m}\nonumber \\
     & +\dfrac{\hbar^2\beta}{2m}\int\mathrm{d}xv_{\sigma}(x)+\frac{\hbar^3\mathcal{I}_{1\mathrm D}\beta^{3/2}}{3\sqrt{2\pi m^3}}+O(\beta^2),
\end{align}
where
\begin{eqnarray}
    N_\sigma &=& \int_{-\infty}^\infty\mathrm{d}x\, n_\sigma(x), \\
    \mathcal{I}_{1\mathrm D} &=& \int_{-\infty}^\infty\mathrm{d}x\, C_{1\mathrm D}(x).
\end{eqnarray}
With the help of the $N$-body Schr\"{o}dinger equation, we find
\begin{eqnarray}
    \sum_\sigma J_\sigma(\beta) &=& \sum_{\nu\sigma}n_{\nu\sigma}e^{-\beta\epsilon_\nu} \nonumber \\
    &=& N-\beta\sum_\sigma\int\mathrm{d}x\left[V(x)n_\sigma(x)-\dfrac{\hbar^2}{2m}v_{\sigma}(x)\right]\nonumber \\
    & &-\dfrac{\hbar^2a_{1\mathrm{D}}\mathcal{I}_{1\mathrm{D}}\beta}{2m}+\frac{2\hbar^3\beta^{3/2}\mathcal{I}_{1\mathrm{D}}}{3\sqrt{2\pi m^3}} \nonumber \\
    &=& N-E\beta-\dfrac{\hbar^2a_{1\mathrm{D}}\mathcal{I}_{1\mathrm{D}}\beta}{2m} \notag \\
    & &+\frac{2\hbar^3\mathcal{I}_{1\mathrm{D}}\beta^{3/2}}{3\sqrt{2\pi m^3}}+O(\beta^2).
\end{eqnarray}
Applying $\frac{\mathrm{d}}{\mathrm{d}\beta}$ to the above expansion and taking $\beta\rightarrow0$, we get
the energy functional shown in \Eq{eq:energyfunc-1D}.
Clearly, the energy functional only gains an extra finite shift, $-\frac{\hbar^2a_{1\mathrm D}\mathcal{I}_{1\mathrm D}}{2m}$, due to the interaction. 
In 1D, the energy theorem is \cite{tan-relation-1dfermi1}
\begin{equation}\label{eq:energytheorem1D}
    E=\sum_\sigma\int\frac{\mathrm{d}k}{2\pi}\frac{\hbar^2k^2}{2m}n_\sigma(k)-\frac{\hbar^2 a_{1\mathrm{D}}\mathcal{I}_{1\mathrm D}}{2m}+\expval{V}.
\end{equation}
If there is no external potential, namely if $V\equiv0$,
the energy functional in \Eq{eq:energyfunc-1D} reduces to this energy theorem.

We also calculate the asymptotics of $\rho(\epsilon)$ and $n_{\nu\sigma}$ in 1D, and the results are
\Eq{eq:rho1Dtail} and
\begin{equation}
    n_{\nu\sigma} = \frac{1}{k_\nu^4}\int\text{d}x\, C_{1\mathrm{D}}(x)|\phi_\nu(x)|^2+O(\epsilon_\nu^{-5/2}),\label{eq:highenergystates1D}
\end{equation}
where $k_\nu=\sqrt{2m\epsilon_\nu}/\hbar$.

\section{Summary and discussion}\label{sec6:summary}
In this work, we have generalized the universal energy functional for trapped two-component Fermi gases from 3D to lower spatial dimensions. We have shown that in lower dimensions the total energy of two-component fermions with zero-range interaction trapped in any smooth potential can be expressed as linear functionals of the occupation probabilities of one-particle energy eigenstates, just like in 3D. We first calculated the one-particle density matrix of two-component fermions by using the Bethe-Peierls boundary conditions. We have also calculated the asymptotic formulas of the occupation probabilities of single-particle levels at high energy. 

The energy functional [\Eq{eq:energyfunc-2D} in 2D, or \Eq{eq:energyfunc-1D} in 1D] is a universal functional, and it holds for all finite-energy states, i.e. both few-body and many-body states, both pure and mixed states, 
both zero-temperature and finite-temperature states. 
It will be important to understand the nontrivial constraints on the occupation probabilities of the single-particle levels,
because such understanding will enable one to determine the many-body ground state energies
by minimizing the energy functional in the presence of such constraints.
One might be able to generalize the energy functional to multi-component fermions, to fermions with unequal masses, and to bosons. 
Future experiments might be able to measure both the occupation probabilities of single-particle levels and the many-body energies of the systems
we have studied. Such experiments should verify the energy functionals that we have derived.

\begin{acknowledgments}
This work was supported by the National Key R$\&$D Program of China (Grants No. 2019YFA0308403 and No. 2021YFA1400902).
\end{acknowledgments}

\bibliography{Universal_Energy_Functionals_for_Trapped_Fermi_Gases_in_Low_Dimensions} 

\end{document}